\documentclass[11pt]{article}
\usepackage[a4paper,margin=25mm]{geometry}
\usepackage{amsmath,amssymb,amsthm,booktabs,graphicx,hyperref}
\usepackage[T1]{fontenc}
\hypersetup{colorlinks=true,linkcolor=blue,citecolor=blue,urlcolor=blue}
\newtheorem{theorem}{Theorem}[section]

\newcommand{\norm}[1]{\left\|#1\right\|}

\title{Controlled preparation and unconditional transport on a fluctuating quantum graph}
\author{Xianbin Wang\\{\small Key Laboratory of Adolescent Cyberpsychology and Behavior}\\{\small (Central China Normal University), Ministry of Education}\\\texttt{chaishen10000@163.com}}
\date{10 November 2026}
\begin{document}
\maketitle
\begin{abstract}
Lieb--Robinson bounds limit the spread of quantum information, but do not establish that a prescribed input can be prepared and transmitted with appreciable probability. For quantum graphs with fluctuating edges, connecting spectral propagation to unconditional readout also requires control of graph coherence, finite boundaries, and preparation errors. We address this problem in a one-probe model on a prescribed square candidate lattice, combining a generalized Rokhsar--Kivelson parent, spectral reduction, adiabatic preparation, and local dynamical estimates. Assuming a supplied coherent graph ground state, we construct a finite-source wave packet and a hopping ramp that populate an isolated branch with strictly positive group velocity. We prove an unconditional arrival probability of at least $5/8$ in a specified receiver region of a sufficiently large finite open square. Explicit bounds on local joint-state and Hamiltonian errors retain a lower probability of $9/16$, without requiring global many-body state closeness or postselection on graph measurements. Exact arithmetic certificates and a separate small-system numerical diagnostic support the quantitative analysis. The construction provides a model-specific benchmark for transport and readout in fluctuating quantum networks, linking a propagation mechanism to an observable lower bound and identifying the local error budgets required to preserve it. It distinguishes suppression of unwanted influence from guaranteed signal delivery, a useful separation for quantum-network and emergent-geometry models. The resource estimates are conservative, graph-state supply is assumed, and the result does not establish background-free geometry or gravitational dynamics.
\end{abstract}

\section{Introduction}
Locality is an operational property of dynamics: an observable at a receiver must respond in a controlled way to an input. Lieb--Robinson bounds quantify the suppression of influence outside a propagation region in local quantum lattice systems \cite{lr,nsy}. Such bounds are upper bounds. They do not establish that a particular input has appreciable arrival probability, that a desired propagating state can be prepared, or that an infinite-volume statement survives finite boundaries and imperfect controls.

These distinctions are especially relevant when interaction channels are quantum degrees of freedom. Quantum graph models, including Quantum Graphity \cite{graphity}, motivate studying matter coupled to changing adjacency. The present work addresses a restricted question: can coherent edge fluctuations coexist with a specified preparation protocol and a positive, unconditional probe-arrival guarantee? We prescribe a square candidate geometry and allow its edge occupations to fluctuate. Thus finite-range locality is built into the candidate support; we do not derive that support from a background-free theory.

The construction combines established tools: generalized Rokhsar--Kivelson (RK) or stochastic-matrix-form parents \cite{rk}, Dobrushin comparison and spectral-gap estimates \cite{dobrushin,wu}, the Feshbach--Schur map \cite{schur}, quantitative adiabatic reasoning \cite{adiabatic}, and quasi-local dynamics \cite{nsy}. Our model-specific result is their compatible use in one fully specified preparation and readout problem, with constants, a finite-source wavefunction, a receiver independent of the unknown exact velocity, and a finite-box error budget. We do not claim that these component methods, or quantum propagation on fluctuating graphs in general, are new.

There are three distinct statements. First, a gapped coherent graph state supports an isolated dispersive probe branch. Second, a product input and a controlled ramp produce a packet that reaches a growing receiver region with positive probability. Third, this probability is stable under quantified local initial-state and Hamiltonian errors. None of these statements requires conditioning on graph measurements. Conversely, none alone proves that a measured macroscopic connected component persists throughout the protocol. We keep that geometric claim outside the theorem.

\section{Model, protocol and principal result}\label{sec:model}
Let $V_\Lambda\subset\mathbb Z^2$ be a finite open square and $E_\Lambda$ its internal nearest-neighbor edges. Edge $e$ carries a qubit with occupation $n_e$. Distinct edges satisfy $e\sim f$ if they share a vertex, and unordered pairs are counted once. Define
\begin{equation}
 w_\Lambda(n)=4^{\sum_en_e+\sum_{\{e,f\}:e\sim f}n_en_f},\qquad
 |\Omega_\Lambda\rangle=Z_\Lambda^{-1/2}\sum_n\sqrt{w_\Lambda(n)}|n\rangle.
 \label{eq:state}
\end{equation}
The local parent and hopping are
\begin{align}
 H_{G,\Lambda}&=\sum_{e\in E_\Lambda}h_e,\quad
 h_e=\frac1{1+r_e}\begin{pmatrix}r_e&-\sqrt{r_e}\\-\sqrt{r_e}&1\end{pmatrix}_e,
 \quad r_e=4^{1+\sum_{f\sim e}n_f},\label{eq:parent}\\
 T_\Lambda&=-\sum_{e=xy\in E_\Lambda}n_e
 (\sigma_x^+\sigma_y^-+\sigma_y^+\sigma_x^-).
 \label{eq:hopping}
\end{align}
The neighboring occupations are diagonal controls in $h_e$. Probe qubits at vertices give a local tensor-product description; the exactly one-probe sector is invariant. In that sector $\norm{T_\Lambda}\le4$. Every $h_e$ is a projection and annihilates $\Omega_\Lambda$. Connectedness of the single-edge-flip configuration graph fixes the positive amplitude ratios, hence this finite graph ground state is unique.

Starting from $\Omega_\Lambda\otimes\phi_{R_0}$, ramp
\begin{equation}
 H_\Lambda(u)=H_{G,\Lambda}+s(u)T_\Lambda,\qquad
 s(u)=\min\{u/\tau,1\},\qquad 0\le u\le T=\tau+t.
 \label{eq:ramp}
\end{equation}
The ramp requires external work. Energy is conserved only after it ends. Units are $\hbar=1$ and graph-projector coefficient one. For a receiver set $S$, the effect
\begin{equation}
 O_S=I-\prod_{x\in S}(I-n_x^{\rm probe})
 \label{eq:receiver}
\end{equation}
has norm one on the full spin algebra and equals the arrival projector in the one-probe sector. Its expectation traces over all graph outcomes.

\begin{theorem}[Specified finite-system arrival]\label{thm:main}
For the model \eqref{eq:state}--\eqref{eq:ramp}, there is the explicit finite-source wavefunction of Section~\ref{sec:packet}, a finite open square, and the predetermined receiver of \eqref{eq:rectangle}, such that
\begin{equation}
 P_\Lambda(T)=\langle O_S\rangle_T\ge5/8.
\end{equation}
Sufficient parameters are listed in Table~\ref{tab:resources}. The statement assumes supply of $\Omega_\Lambda$ and the specified probe input, and the ramp \eqref{eq:ramp}. It uses neither energy filtering nor graph postselection. It is a finite-resource existence statement, not an experimentally practical resource estimate.
\end{theorem}

The proof proceeds through Sections~\ref{sec:bulk}--\ref{sec:finite}; Appendix~\ref{app:adia} supplies the adiabatic constants. Section~\ref{sec:robust} replaces exact initial-state and control assumptions by explicit local error conditions. The receiver grows with travel distance and is not a single-site detector.

\section{Coherent bulk graph state and gap}\label{sec:bulk}
The classical conditional opening probability is $4^{1+k}/(1+4^{1+k})$. Changing one neighbor changes it by at most $12/85$. The edge line graph has degree six, so the Dobrushin row and column sums are at most
\begin{equation}
 \alpha=72/85<1,\qquad g=1-\alpha=13/85.
\end{equation}
Uniqueness and convergence of local Gibbs marginals follow \cite{dobrushin}; the corresponding conditional Poincare estimate has constant $g$ \cite{wu}.

It is important not to replace the quantum state by its diagonal distribution. On $\mathcal K=L^2(\mu)$, set $\Omega=1$, represent occupations by multiplication, and represent the Pauli flip by
\begin{equation}
 (X_ef)(n)=\sqrt{\rho_e(n)}f(n^e),\quad
 \rho_e(n)=4^{(1-2n_e)(1+\sum_{f\sim e}n_f)}.
 \label{eq:flip}
\end{equation}
Here $n^e$ flips bit $e$; $\rho_e$ is a local conditional ratio, not a quotient of infinite-configuration point probabilities. Gibbs change of variables proves that $X_e$ is a self-adjoint unitary. The two-flip cocycle ensures that different flips commute, and $X_en_eX_e=1-n_e$. This realizes the local edge-qubit algebra, including off-diagonal expectations.

In this representation $h_e=I-\mathbb E_e$, with $\mathbb E_e$ the conditional expectation given all other bits. On cylinder functions define $K_0=\sum_e(I-\mathbb E_e)$. If $f$ depends on $m$ edges, support counting of individual product terms gives
\begin{equation}
 \norm{K_0^nf}\le\norm f\prod_{j=0}^{n-1}(m+6j).
\end{equation}
Thus cylinder functions are analytic vectors for $|u|<1/6$, and Nelson's theorem \cite{nelson} makes $K_0$ essentially self-adjoint. Conditional finite-volume Poincare inequalities, reverse martingale convergence, and tail triviality of the unique Gibbs measure imply
\begin{equation}
 H_G=\overline{K_0}\ge g(I-|\Omega\rangle\langle\Omega|).
 \label{eq:bulk_gap}
\end{equation}
Appendix~\ref{app:bulk} gives the domain and local-state arguments. Square symmetries are unitarily implemented, fix $\Omega$, and commute with $H_G$. The gap and ground-vector uniqueness refer to this representation.

\section{Certified dispersive branch}\label{sec:band}
The relative-coordinate transform $(U\psi)_x=\mathsf T_{-x}\psi_x$, with $\mathsf T_x$ the graph translations, preserves position probabilities. Fourier transformation yields
\begin{equation}
 H_s(k)=H_G+sT(k),\quad T(k)=-\sum_{e=\pm e_1,\pm e_2}e^{ik\cdot e}n_{0e}\mathsf T_e.
\end{equation}
One has $\norm T\le4$ and $\norm{\partial_{k_j}T}\le2$. Let $P_0=|\Omega\rangle\langle\Omega|$, $Q_0=1-P_0$, $p=\langle n_e\rangle$, and
\begin{equation}
 a=-2p(\cos k_1+\cos k_2),\quad B=Q_0T\Omega,\quad D=Q_0(H_G+T)Q_0.
\end{equation}
Exact integer enumeration of a 17-edge open $4\times3$ rectangle gives its central-edge closing probability
\begin{equation}
 q_*=\frac{45990846799967188519203511}{750185649985219117820588975027}<q:=1/15000.
 \label{eq:q}
\end{equation}
Attractive heat-bath coupling compares zero exterior bits to arbitrary exterior configurations. Averaging those conditional laws in the bulk and using translation symmetry gives $1-p\le q_*<q$. Appendix~\ref{app:exact} describes the finite sum. This is a proved comparison, not a finite-size extrapolation of a quantum spectrum.

Choose
\begin{equation}
 k_0=(2/25,0),\quad \rho=10^{-5},\quad
 W_k=\{|k_1-2/25|\le\rho,\ |k_2|\le\rho\}.
\end{equation}
On $W_k$, $a(k)\le-3.99333215990668<-3.99$. Since $Q_0H_sQ_0\ge g-4s$ and the trial vector gives $E_s\le sa$, the codimension-one variational principle gives one simple eigenvalue below $g-4s$. The remaining spectrum lies above that threshold and
\begin{equation}
 \operatorname{gap}_s(k)\ge g-s[4+a(k)]\ge\delta:=243/1700.
 \label{eq:fibergap}
\end{equation}
Bounded analytic perturbations of the common-domain operator give a smooth branch and projections. No global spectral gap of the direct integral over $k$ is assumed.

At $s=1$ the scalar Schur equation \cite{schur} is
\begin{equation}
 E=a+\Sigma(k,E),\quad \Sigma(k,z)=\langle B,(z-D)^{-1}B\rangle.
\end{equation}
For $z\le-3.99$, $\norm B\le4\sqrt q$, $\norm{B_{k_j}}\le2\sqrt q$ and $\norm{D_{k_j}}\le2$ imply
\begin{equation}
 |\Sigma_{k_j}|\le\epsilon_D:=16q/\delta+32q/\delta^2,
 \qquad1\le1-\Sigma_z\le1+16q/\delta^2.
\end{equation}
Differentiation gives $E_{k_j}=(a_{k_j}+\Sigma_{k_j})/(1-\Sigma_z)$. Using $\sin x\ge x-x^3/6$ at $x=2/25$ certifies
\begin{equation}
 0.0455670986857\ldots\le v:=E_{k_1}(k_0)\le0.2718727384602\ldots,
 \quad \boxed{1/25<v<3/10.}
 \label{eq:v}
\end{equation}
Reflection gives $E_{k_2}(k_0)=0$. For real unit $u$, $\norm{H_u}\le2\sqrt2$ and $\norm{H_{uu}}\le2$. The reduced-resolvent second derivative formula bounds
\begin{equation}
 \norm{\nabla^2E}\le2+16/\delta=27686/243,
 \quad |\nabla E(k)-(v,0)|\le(27686/243)\sqrt2\rho<1/400\le v/16.
 \label{eq:spread}
\end{equation}

\section{Input, preparation and unconditional bulk arrival}\label{sec:packet}
Use normalized Haar measure $d^2k/(2\pi)^2$ and the explicit envelope
\begin{equation}
 h(u)=\begin{cases}\sqrt{315/256}(1-u^2)^2,&|u|\le1,\\0,&|u|>1,\end{cases}
 \quad f(k)=\frac{2\pi}{\rho}h\left(\frac{k_1-2/25}{\rho}\right)h\left(\frac{k_2}{\rho}\right).
\end{equation}
It is normalized and belongs to $H^1$, with $C_f^2=\norm{\nabla f}^2=6/\rho^2$, hence $C_f\le250000$. $H^1$ regularity suffices for all position-second-moment estimates below. Let $\phi$ be its inverse Fourier transform and define the predetermined finite-source input
\begin{equation}
 \phi_{R_0}=\frac{\mathbf1_{|x|\le R_0}\phi}{\norm{\mathbf1_{|x|\le R_0}\phi}},
 \qquad\norm{\phi-\phi_{R_0}}\le\frac{3C_f}{2R_0}.
 \label{eq:source}
\end{equation}
Specifying this vector does not perform postselection. The relative transform leaves the product input unchanged because translations fix $\Omega$.

Quantitative adiabatic reasoning \cite{adiabatic}, with the bounded-commutator proof in Appendix~\ref{app:adia}, gives a uniform vector error $K/\tau$, where
\begin{equation}
 K=16/\delta^2+576/\delta^3=315680480000/1594323.
\end{equation}
The ideal adiabatic output is
\begin{equation}
 \widehat\psi_A(k)=f(k)e^{-i\tau\int_0^1E_s(k)ds}\,W(k,1)\Omega,
\end{equation}
where $W$ is Kato parallel transport. Both dynamical and geometric phases are retained. Their derivative estimates give
\begin{equation}
 C_\tau:=\norm{\nabla\widehat\psi_A}\le C_f+3\tau/2+G,
 \quad G=480/\delta^2+12/\delta=464052400/19683.
\end{equation}
The phase bound includes motion during preparation; we do not reset the packet center. After flight time $t$ under $H_1$,
\begin{equation}
 \norm{(X/t-(v,0))\psi_A(t)}\le v/16+C_\tau/t.
\end{equation}
For $t\ge16C_\tau/v$, Chebyshev's inequality gives probability $3/4$ in the ball centered at $(vt,0)$ of radius $vt/4$. Pure-state trace distance is at most vector distance; unitarity preserves the source error. Thus actual bulk probability in this ball is at least
\begin{equation}
 3/4-K/\tau-3C_f/(2R_0)\ge11/16
\end{equation}
when each subtracted term is at most $1/32$. Source truncation introduces other momenta, but its norm error is propagated exactly; no gap outside $W_k$ is needed.

The exact $v$ is not computed. The preselected rectangle
\begin{equation}
 S_t=\{x\in\mathbb Z^2:3t/100\le x_1\le3t/8,\quad |x_2|\le3t/40\}
 \label{eq:rectangle}
\end{equation}
contains that ball for every $v\in[1/25,3/10]$, and therefore inherits its bound.

\section{Finite boundaries and the quantum initial state}\label{sec:finite}
Two metrics must be distinguished. The incidence metric $d_I$ has a site for every vertex probe and edge qubit, with edge sites joined to endpoints. The edge line metric $d_E$ joins edge qubits sharing a vertex. A parent term has incidence diameter at most four and at most seven sites; a hopping term has three sites. Uniformly in time,
\begin{equation}
 \sup_x\sum_{Z\ni x}\norm{h_Z}\le8,\qquad
 \sup_x\sum_{Z\ni x}|Z|\norm{h_Z}\le52.
 \label{eq:strength}
\end{equation}
These full-spin-algebra bounds avoid the extensive norm of $H_G$.

For a graph-edge region $A$, let $B$ be the set of exterior edges incident to $A$. If $A\cup B\subset E_\Lambda$, finite and bulk RK reduced states obey
\begin{equation}
 \rho_A^j=\sum_b(\mu_j)_B(b)|\psi_A^b\rangle\langle\psi_A^b|,
 \quad |\psi_A^b\rangle=\sum_a\sqrt{\mu(a\mid b)}|a\rangle,\quad j\in\{\Lambda,\infty\}.
 \label{eq:coherent}
\end{equation}
Their conditional coherent vectors are identical; only the boundary mixing weights differ. This preserves all off-diagonal entries. If $d$ is the edge-line distance of $B$ from the layer of finite edges adjacent to omitted exterior edges, Dobrushin comparison gives
\begin{equation}
 D(\rho_A^\Lambda,\rho_A^\infty)\le\operatorname{TV}((\mu_\Lambda)_B,\mu_B)
 \le |B|\alpha^d/(1-\alpha).
 \label{eq:initialcomparison}
\end{equation}

Let $W_r=\{z:d_I(z,S_t)\le r\}$, and use the common Hamiltonian consisting of terms wholly supported in $W_r$. Place $W_r$ at least eight incidence steps from the boundary-modified parents, and include the source in the box. Standard interaction-path iteration \cite{nsy}, detailed in Appendix~\ref{app:lr}, bounds each dynamics-to-window difference by $16|S_t|T e^{300T-(r-4)/4}$. The factor $T=\tau+t$ includes the ramp. Combining the two dynamics differences with \eqref{eq:initialcomparison} yields
\begin{equation}
 \boxed{|P_\Lambda(T)-P_\infty(T)|\le
 32|S_t|T e^{300T-(r-4)/4}+|B|\alpha^d/(1-\alpha).}
 \label{eq:transfer}
\end{equation}
Choosing each error at most $1/32$ proves Theorem~\ref{thm:main}. The bound concerns a family of increasing boxes as time grows, not arbitrarily long directed propagation in one fixed box.

\begin{table}[t]
\centering\begin{tabular}{lr}\toprule
Quantity & Sufficient value\\\midrule
Momentum center & $(0.08,0)$\\
Momentum half-width & $10^{-5}$\\
Source radius $R_0$ & $12\,000\,000$\\
Ramp time $\tau$ & $6\,336\,091$\\
Flight time $t$ & $3\,911\,085\,122$\\
Total time $T$ & $3\,917\,421\,213$\\
Incidence buffer radius $r$ & $4\,700\,905\,456\,016$\\
Static edge-line depth $d$ & $652$\\
Square half-side $L$ & $4\,702\,384\,113\,609$\\\bottomrule
\end{tabular}
\caption{An exact-arithmetic sufficient certificate in dimensionless units. The box has vertices $[-L,L]^2\cap\mathbb Z^2$. These deliberately loose resources establish finiteness, not practicality.}\label{tab:resources}
\end{table}

\section{Local initial-state and control errors}\label{sec:robust}
We now compare the finite ideal protocol with an imperfect initial density matrix $\widetilde\rho$ and Hamiltonian $\widetilde H(u)=H(u)+V(u)$, $V=\sum_ZV_Z$, on the same finite box. Terms are bounded and measurable. To interpret the effect as arrival of the injected probe, both states are supported in the one-probe sector and the actual Hamiltonian preserves probe number. Without this restriction the estimate remains one about detection, which could include unwanted probe creation.

Let $\rho$ be the ideal joint initial state, $W_r$ the receiver window, and
\begin{equation}
 \eta_W=D(\widetilde\rho_{W_r},\rho_{W_r}),\qquad
 E_W=32|S|T e^{300T-(r-4)/4}.
\end{equation}
$\eta_W$ is a \emph{joint} reduced-state distance, including all graph--probe correlations in the window. Agreement of one-site densities, small defect density, or separate graph and probe marginals does not imply this condition.

For the ideal interaction set put
\begin{equation}
 K_{xy}=\sum_{Z\ni x,y}\sup_u\norm{h_Z(u)},\qquad
 M_{SZ}(a)=\sum_{x\in S,y\in Z}[e^{2aK}]_{xy}.
\end{equation}
Define the observable-specific control budget
\begin{equation}
 E_V=2\int_0^T\sum_Z\norm{V_Z(u)}
             \min\{1,M_{SZ}(T-u)\}\,du.
 \label{eq:controlbudget}
\end{equation}
This is finite in a finite box and has useful volume-independent bounds for local error families, as specified below.

\begin{theorem}[Local robustness of readout]\label{thm:robust}
For an actual receiver effect $0\le\widetilde O_S\le I$ with $\norm{\widetilde O_S-O_S}\le\eta_M$,
\begin{equation}
 \boxed{|\widetilde P(T)-P_\Lambda(T)|\le\eta_W+E_W+E_V+\eta_M.}
 \label{eq:robust}
\end{equation}
No global distance between the initial states is assumed. All terms can be capped by the trivial probability bound when appropriate.
\end{theorem}
\begin{proof}
First compare the two initial states under ideal dynamics. Approximate the evolved effect by the effect evolved with terms wholly within $W_r$. Each expectation changes by at most $16|S|T e^{300T-(r-4)/4}$. The window effect lies between zero and one, so its state difference is at most $\eta_W$. Next compare actual and ideal evolutions on the same actual initial state. The variation-of-propagators identity bounds the operator difference by the integral of $\norm{[V(u),\mathcal U_H(T,u)(O_S)]}$. The intervening actual propagators are unitary and disappear from the norm estimate. Apply both the trivial commutator bound and the ideal interaction-path bound termwise to obtain \eqref{eq:controlbudget}. Finally the effect perturbation contributes at most $\eta_M$.
\end{proof}

The result does not require proving a new spectral gap for the perturbed Hamiltonian. It controls a finite-time observable directly. It does not treat dissipative or measurement-conditioned evolution.

For a coarse bound, choose a radius $R$, split error terms according to whether their support meets $W_R$, and let $\lambda_{\rm near}(u)$ and $\lambda_{\rm far}(u)$ bound their respective per-site summed norms. Then
\begin{equation}
 E_V\le2\int_0^T\left[|W_R|\lambda_{\rm near}(u)
 +|S|\lambda_{\rm far}(u)e^{300(T-u)-R/4}\right]du.
 \label{eq:split}
\end{equation}
For the near sum every term is charged to at least one site of $W_R$; for the far sum every endpoint is at distance greater than $R$, so the weighted kernel tail applies. The bound depends on the relevant space--time region, not the full box volume. Its constants need not be useful for uniform noise over extremely long protocols.

A useful special case is coherent ramp calibration error $V(u)=\delta s(u)T_\Lambda$. In the one-probe sector $\norm{T_\Lambda}\le4$, so
\begin{equation}
 E_V\le8\int_0^T|\delta s(u)|\,du.
 \label{eq:ramp_error}
\end{equation}
This avoids multiplying the calibration error by the number of edges. If $\eta_W\le1/64$, $E_W\le1/64$, $E_V\le1/32$, and $\eta_M=0$, the ideal $5/8$ guarantee becomes
\begin{equation}
 \widetilde P(T)\ge9/16.
 \label{eq:robustlower}
\end{equation}
The window can be enlarged slightly, retaining a sufficiently large box, to meet the tighter $E_W$ budget. For ramp-only error, $\int|\delta s|\le1/256$ suffices. With Table~\ref{tab:resources}, a constant offset confined to the ramp would require $|\delta s|\le1/(256\tau)\simeq6.17\times10^{-10}$ under this bound. These are sufficient tolerances, not realistic demonstrated specifications.

Local state closeness is an assumption to be certified by preparation analysis. For product inputs it follows from graph and probe local trace-distance bounds by a triangle inequality. More generally a measured or bounded correlation error must also be included. The theorem allows a mixed actual state and arbitrary exterior correlations consistent with the stated joint marginal; it does not itself provide a graph-ground-state preparation algorithm.

\section{Reproducibility and a separate numerical diagnostic}\label{sec:numerics}
The supplied standard-library programs enumerate the 17-edge classical rectangle with exact integer weights and verify every exponent by an independent degree-count formula. Rational calculations check the spectral inequalities, envelope moments, times, boundary budgets and robustness budgets. A two-edge reduced RK state on a $3\times3$ box is checked entrywise, including nonzero off-diagonal entries, against \eqref{eq:coherent}. Support counts are checked on three open-box sizes. These tests verify algebra and arithmetic; the proofs above supply the infinite-volume and propagation arguments.

For a small end-to-end dynamics diagnostic, we also simulate a $2\times3$ open box with seven edge qubits and one probe. The joint dimension is $2^7\times6=768$. We prepare its exact RK graph state and a \emph{single-site} probe at $(0,0)$, ramp for time one, and propagate to total time four. The receiver is $(1,2)$. Two controls apply a graph rotation $e^{0.02iX_e}$ at the source-side edge or a hopping-coefficient offset $0.005$ during the ramp.

\begin{figure}[t]
\centering\includegraphics[width=\linewidth]{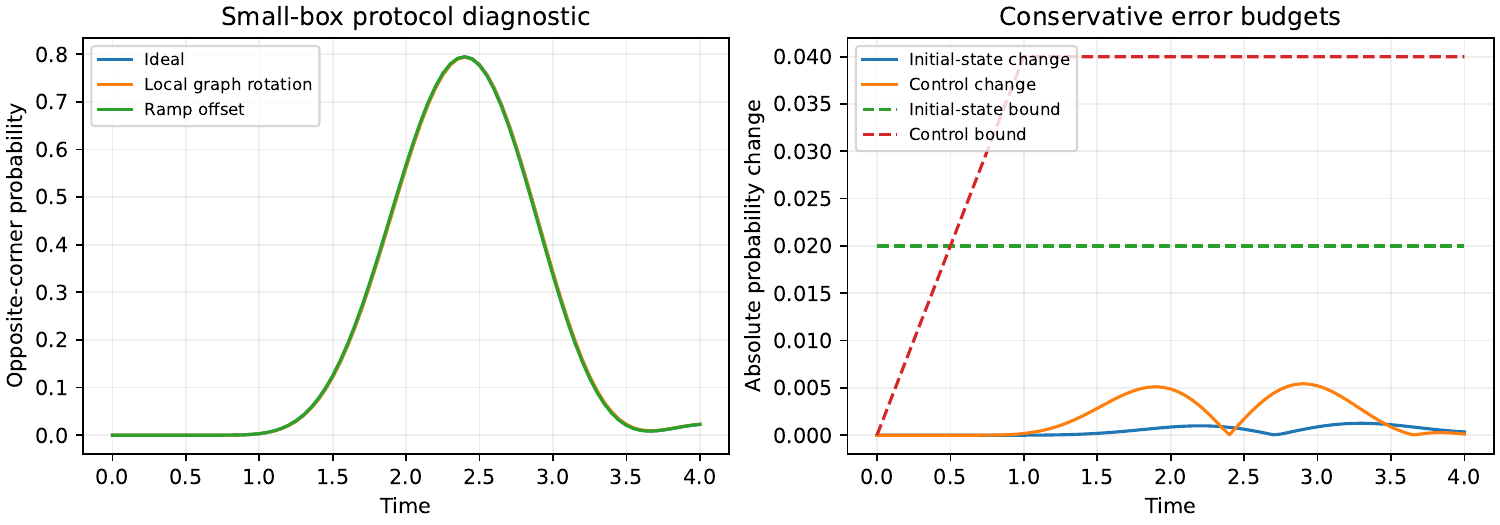}
\caption{A separate 768-dimensional finite-box diagnostic, with a single-site input and short times. Left: unconditional receiver probability for ideal and perturbed protocols. Right: probability changes and conservative error bounds. This is not the explicit macroscopic packet of Theorem~\ref{thm:main}, and does not numerically validate its $5/8$ bound.}\label{fig:diagnostic}
\end{figure}

Piecewise-constant midpoint propagation with steps $0.025$ and $0.0125$ gives a maximum receiver-probability refinement difference \mbox{$3.51\times10^{-7}$} and maximum normalization drift \mbox{$2.38\times10^{-14}$}. Maximum probability changes from the initial rotation and ramp offset are approximately $0.0012644$ and $0.0054428$, respectively, below their conservative bounds $0.02$ and $0.04$. Figure~\ref{fig:diagnostic} shows the full curves. Grid refinement is a floating-point consistency check, not a rigorous time-discretization certificate.

\section{Discussion and scope}
The result provides a sufficient preparation-to-readout construction in a quantum graph with fluctuating edge occupations. The explicit positive group-velocity interval supplies a lower transport statement, while local propagation estimates control unwanted boundary influence. Those are complementary roles. Off-diagonal graph coherence is retained throughout the initial-state comparison and the joint evolution.

There are substantial limitations. The square candidate geometry and spatially local control support are prescribed. Edge occupation fluctuations do not demonstrate emergence of dimension or metric. The source and receiver are large, and Table~\ref{tab:resources} is far beyond practical implementation. The model has one probe; finite-density matter, backreaction from many probes and relativistic limits are outside the proof. Exact or locally accurate graph-state supply is assumed. A probability that the graph has a giant measured component is distinct from the present dynamical arrival bound, and simultaneous preservation of both has not been established here.

The local robustness theorem identifies what a preparation or control analysis must certify, but it does not certify a laboratory. Its general comparison structure is standard quasi-local reasoning; the contribution here is an explicit compatibility check with this model's coherent state, band, ramp and receiver. Further work can improve the constants, construct locally accurate graph-state preparations, and test which parts persist across a class of parents. These are extensions of the stated result, not premises silently used in its proof.

\section*{Data and code availability}
Reproducibility scripts, exact certificates, numerical CSV data and figure-generation code are provided in the ancillary files accompanying this submission. The exact checks use Python's standard library; the numerical diagnostic uses NumPy, SciPy and Matplotlib with versions recorded in its JSON output. No separate public repository or DOI has been assigned.
\section*{Funding and competing interests}
No dedicated funding was received. The author declares no competing interests.

\section*{Use of AI-assisted tools}
OpenAI ChatGPT and Codex were used extensively to assist with mathematical exploration, derivation drafting, code development, computational checks, and manuscript preparation. The supplied scripts document the computational checks; these checks and AI-assisted analysis do not constitute independent mathematical peer review. Responsibility for the content rests with the named author.

\appendix
\section{Bulk representation, domain and gap}\label{app:bulk}
For the unique Gibbs measure $\mu$, conditional change of variables gives
\begin{equation}
 \int F(n^e)\rho_e(n)d\mu(n)=\int F(n)d\mu(n).
\end{equation}
Together with $\rho_e(n)\rho_e(n^e)=1$ this makes \eqref{eq:flip} unitary and self-adjoint. Flips on distinct edges commute because
\begin{equation}
 \rho_e(n)\rho_f(n^e)=\rho_f(n)\rho_e(n^f).
\end{equation}
Diagonal cylinder functions acting on $\Omega=1$ are dense, so the representation is cyclic. If a local matrix $O$ is supported on edge set $A$, its finite RK expectation equals the classical expectation of
\begin{equation}
 f_O(n)=\sum_{m_A}O_{n_A,m_A}
 \sqrt{\frac{w(m_A,n_{A^c})}{w(n_A,n_{A^c})}}.
\end{equation}
Only interactions touching $A$ enter the ratio. Thus $f_O$ is a bounded function of $A$ and its finite neighboring edge set. Classical local convergence establishes convergence of full quantum expectations, not just diagonal observables.

For fixed neighbors with $r=4^{1+k}$, conditional expectation on the two edge values is the matrix
\begin{equation}
 \mathbb E_e=\begin{pmatrix}(1+r)^{-1}&r(1+r)^{-1}\\(1+r)^{-1}&r(1+r)^{-1}\end{pmatrix}.
\end{equation}
It is orthogonal in the weighted conditional inner product. Inserting the weighted flip identifies the local parent with $I-\mathbb E_e$.

The cylinder domain is invariant under $K_0$. In a product of $n$ parent projections acting on a function with $m$ coordinates, only coordinates already in the current term's support can act nontrivially. Each action adds at most six coordinates. There are at most $\prod_{j<n}(m+6j)$ ordered nonzero terms, each with norm at most that of the original vector. Consequently
\begin{equation}
 \sum_{n\ge0}\frac{|u|^n}{n!}\norm{K_0^nf}
 \le\norm f(1-6|u|)^{-m/6},\qquad |u|<1/6.
\end{equation}
Constants are killed by $K_0$. The dense invariant analytic-vector domain gives essential self-adjointness \cite{nelson}.

For any fixed exterior configuration, the conditional measure on a finite edge set has the same influence bound: exterior occupied bits only increase nonnegative fields. Its Poincare inequality \cite{wu} implies, after integrating the exterior,
\begin{equation}
 \mathbb E_\mu\operatorname{Var}(f\mid\Lambda^c)
 \le g^{-1}\sum_e\norm{(I-\mathbb E_e)f}^2.
\end{equation}
Only coordinates in the support of $f$ contribute. Uniqueness of the Gibbs measure implies tail triviality, since conditioning on a nontrivial tail event would give two distinct Gibbs measures. The reverse martingale theorem then sends the left side to $\operatorname{Var}_\mu f$ along exhausting sets. Orthogonality of the projections and graph-norm approximation along the operator core extend the resulting inequality to $H_G$, proving \eqref{eq:bulk_gap}.

Translations and reflections act by pullback of the invariant measure, preserve the cylinder core, and commute with the closure. For local $O$, the identity
\begin{equation}
 [K_0,\pi(O)]=\sum_e[\pi(h_e),\pi(O)]
\end{equation}
has only finitely many nonzero bounded terms. Thus the represented dynamics implements the same local derivation as the finite-range parent. The norm local thermodynamic-limit dynamics supplied by the summable interaction bounds \cite{nsy} agrees with this implementation. Adding the bounded one-probe hopping yields a self-adjoint bulk operator on the graph-generator domain.

\section{Exact occupation and envelope calculations}\label{app:exact}
On the open $4\times3$ vertex rectangle there are 17 candidate edges. Integer summation gives
\begin{align}
 Z&=5251299549896533824744122825189,\\
 Z_{n_{((1,1),(2,1))}=0}&=321935927599770319634424577.
\end{align}
Their ratio reduces to \eqref{eq:q}. Every weight exponent is cross-checked against $\sum_en_e+\sum_v\binom{d_v(n)}2$. Conditional opening probabilities are increasing in all neighbors. Monotone coupling with common uniform random numbers proves that zero-exterior heat-bath dynamics is dominated by any fixed exterior-conditioned dynamics. Finite irreducibility gives comparison of their stationary laws; the bulk conditional expectation then proves the inequality used in Section~\ref{sec:band}.

The elementary integrals
\begin{equation}
 \int_{-1}^1(1-u^2)^4du=256/315,\qquad \int |h'(u)|^2du=3
\end{equation}
give the envelope normalization and second moment. Its inverse Fourier transform can be evaluated one position at a time:
\begin{equation}
 \phi_x=\frac{\rho}{2\pi}e^{2ix_1/25}\widehat h(\rho x_1)\widehat h(\rho x_2),
 \quad \widehat h(z)=\int_{-1}^1h(u)e^{izu}du.
\end{equation}
For $z\ne0$,
\begin{equation}
 \widehat h(z)=16\sqrt{315/256}\,
 \frac{(3-z^2)\sin z-3z\cos z}{z^5},\quad
 \widehat h(0)=\frac{16}{15}\sqrt{315/256}.
\end{equation}
Near zero a convergent power series avoids cancellation. The finite source retains norm squared at least $2399/2400$. Its normalization is a specified finite sum, not one enumerated by the supplied software at the enormous certified radius.

\section{Adiabatic constants and momentum-dependent phases}\label{app:adia}
Suppress $k$ and set $P=P_s(k)$, $Q=1-P$, $R=Q(H_s-E_s)^{-1}Q$. A prime denotes $s$ differentiation. The gap \eqref{eq:fibergap} and $M=\norm{H_s'}\le4$ yield
\begin{align}
 P'&=-RH'P-PH'R,&\norm{P'}&\le2M/\delta,\\
 R'&=-R(H'-E')R-RP'-P'R,&\norm{R'}&\le6M/\delta^2.
\end{align}
Let $C=[P',P]$ and $X=RP'P+PP'R=-R^2H'P-PH'R^2$. Then $[H,X]=C$, with bounded extension, and since $H''=0$,
\begin{equation}
 \norm C\le4M/\delta,\quad\norm X\le2M/\delta^2,
 \quad\norm{X'}\le28M^2/\delta^3.
\end{equation}
For propagators $U'=-i\tau HU$ and $U_A'=(-i\tau H+C)U_A$, differentiation of $U^*XU_A$ and integration by parts give
\begin{equation}
 \norm{U(1)-U_A(1)}\le\tau^{-1}
 \left(2\sup_s\norm X+\int_0^1[\norm{X'}+\norm X\norm C]ds\right)\le K/\tau.
\end{equation}
This is the usual bounded-commutator adiabatic mechanism \cite{adiabatic}. Here no finite global norm of $H_G$ is used. Common-domain bounded perturbations define both propagators; the resolvent maps into that domain, so the integration identity holds first there and extends by density. $U_A$ intertwines $P_0$ and $P_s$. Uniformity in momentum allows integration over a packet without a global direct-integral gap.

Kato transport satisfies $\partial_sW=C_sW$. Bounds on momentum derivatives are
\begin{align}
 \norm{P_{k_j}}&\le4/\delta,&\norm{R_{k_j}}&\le12/\delta^2,\\
 \norm{P'_{k_j}}&\le128/\delta^2+4/\delta,&
 \norm{C_{k_j}}&\le320/\delta^2+8/\delta.
\end{align}
The last line follows by differentiating $[P',P]$. Unitarity bounds $\norm{W_{k_j}(1)}$ by the integral of $\norm{C_{k_j}}$. Also $|(E_s)_{k_j}|\le2s$. Taking the two-coordinate norm and using $\sqrt2\le3/2$ proves
\begin{equation}
 \norm{\nabla\widehat\psi_A}\le C_f+3\tau/2+480/\delta^2+12/\delta.
\end{equation}
For finite-source truncation, let $a=\norm{\mathbf1_{|x|>R_0}\phi}^2\le C_f^2/R_0^2$. Then $\norm{\phi-\phi_{R_0}}^2=2(1-\sqrt{1-a})\le2a$, proving \eqref{eq:source}. Partial trace and effects contract trace distance, so all preparation errors transfer to the unconditional arrival probability.

\section{Interaction paths and finite-box comparison}\label{app:lr}
For $K$ in Section~\ref{sec:robust}, repeated Duhamel commutator iteration gives
\begin{equation}
 \norm{[\mathcal U_H(u,v)(A_S),B_Y]}
 \le2\norm{A_S}\norm{B_Y}\sum_{x\in S,y\in Y}[e^{2|u-v|K}]_{xy}.
 \label{eq:kernel_app}
\end{equation}
The iteration is over chains of interaction supports, consecutive supports intersecting. Choosing a shared site for every intersection embeds each support chain into a site path; summing all such choices overcounts it. The initial intersection term is bounded by $2\norm{A_S}\norm{B_Y}|S\cap Y|$. The $n$-th ordered time integral is at most $|u-v|^n/n!$, which yields the matrix exponential. Time-dependent coefficients are controlled by their fixed supportwise norm majorants.

Every term has diameter at most four and \eqref{eq:strength} gives row sum at most 52. Along a path, the triangle inequality multiplies exponential distance weights, so
\begin{equation}
 \sup_x\sum_y e^{\mu d_I(x,y)}[e^{2aK}]_{xy}
 \le e^{104e^{4\mu}a}.
\end{equation}
With $\mu=1/4$, use $104e<300$. For dynamics truncated to $W_r$, only crossing terms enter the variation identity. Each of their sites is at distance at least $r-4$ from $S$. Summing crossing strengths at a fixed site costs at most eight, the commutator gives two, and integration over duration $T$ gives
\begin{equation}
 \norm{\mathcal U_H(T,0)(O_S)-\mathcal U_{W_r}(T,0)(O_S)}
 \le16|S|T e^{300T-(r-4)/4}.
\end{equation}
Summability justifies infinite-volume passage. Boundary-modified finite-box parents have disjoint support from $W_r$ when the stated eight-step margin holds, so the same interior dynamics works on both sides.

For \eqref{eq:coherent}, trace the positive-amplitude RK pure state over the exterior. For each fixed exterior configuration, its internal amplitude vector is the square root of the conditional distribution; the Gibbs Markov property makes it depend only on $b=n_B$. Averaging gives the mixture identity. Contractivity of the classical-to-quantum map yields the trace-distance bound. A Dobrushin comparison coupling gives disagreement probabilities $q\le Cq+b$ with $\norm C_\infty\le\alpha$ and $b\le1$ supported in the boundary layer. An edge at distance $d$ has disagreement at most $\sum_{m\ge d}\alpha^m$; a union bound over $B$ proves \eqref{eq:initialcomparison}.

To certify the resources without enumerating a huge box, take
\begin{align}
 Q&=\left\lceil\log_2(1024|S|\max\{T,1\})\right\rceil,\qquad
 r=\left\lceil4+1200T+4Q\right\rceil,\\
 |B|&\le6|S|[1+2r(r+1)],\qquad
 d\ge\left\lceil\frac{\log(32|B|/(1-\alpha))}{\log(1/\alpha)}\right\rceil.
\end{align}
The cardinality bound embeds incidence sites into doubled integer coordinates. These choices make each error in \eqref{eq:transfer} at most $1/32$. For \eqref{eq:robustlower}, replace 1024 by 2048 to make $E_W\le1/64$, enlarging the box if necessary. No assumption about global wavefunction fidelity enters this operation.

\section{Verification boundary and dependency audit}
The argument uses the same parameters $a=b=4$ and $J=1$ throughout. It distinguishes the full spin algebra, used for locality, from the one-probe invariant sector, used for the norm-four hopping and arrival interpretation. Bulk spectral gaps are fiberwise; the packet is $H^1$; ramp phases are included; total boundary time is $\tau+t$; $d_I$ and $d_E$ are distinct; the new momentum patch has its own gap check; and finite RK local comparison retains off-diagonal coherence. These interfaces are recorded in the accompanying audit file.

The exact certificates do not constitute a formal proof-assistant verification or independent mathematical peer review. The small-box dynamics checks do not certify the large-box probability theorem. The local-error theorem assumes a bound on a full joint reduced state, not a list of single-site moments. No global geometric event is inferred from local trace-distance closeness. All archived preliminary supplements are provenance rather than additional unstated hypotheses of this manuscript.


\clearpage
\begin{thebibliography}{10}
\bibitem{lr} B. Nachtergaele and R. Sims, Lieb--Robinson bounds and the exponential clustering theorem, Commun. Math. Phys. \textbf{265}, 119--130 (2006), \href{https://arxiv.org/abs/math-ph/0506030}{arXiv:math-ph/0506030}.
\bibitem{nsy} B. Nachtergaele, R. Sims and A. Young, Quasi-Locality Bounds for Quantum Lattice Systems. Part I, J. Math. Phys. \textbf{60}, 061101 (2019), \href{https://arxiv.org/abs/1810.02428}{arXiv:1810.02428}.
\bibitem{graphity} T. Konopka, F. Markopoulou and S. Severini, Quantum Graphity: A Model of Emergent Locality, Phys. Rev. D \textbf{77}, 104029 (2008), \href{https://arxiv.org/abs/0801.0861}{arXiv:0801.0861}.
\bibitem{rk} C. Castelnovo, C. Chamon, C. Mudry and P. Pujol, From quantum mechanics to classical statistical physics: generalized Rokhsar--Kivelson Hamiltonians and the stochastic matrix form decomposition, Ann. Phys. \textbf{318}, 316--344 (2005), \href{https://arxiv.org/abs/cond-mat/0502068}{arXiv:cond-mat/0502068}.
\bibitem{dobrushin} T. C. Dorlas and B. Savoie, On the Wasserstein distance and the Dobrushin uniqueness theorem, \href{https://arxiv.org/abs/2305.19371}{arXiv:2305.19371}.
\bibitem{wu} L. Wu, Poincare and transportation inequalities for Gibbs measures under the Dobrushin uniqueness condition, Ann. Probab. \textbf{34}, 1960--1989 (2006), \href{https://arxiv.org/abs/math/0611635}{arXiv:math/0611635}.
\bibitem{nelson} E. Nelson, Analytic vectors, Ann. Math. \textbf{70}, 572--615 (1959), \href{https://doi.org/10.2307/1970331}{doi:10.2307/1970331}.
\bibitem{schur} G. Dusson, I. M. Sigal and B. Stamm, The Feshbach--Schur map and perturbation theory (2021), \href{https://arxiv.org/abs/2105.02058}{arXiv:2105.02058}.
\bibitem{adiabatic} S. Jansen, M.-B. Ruskai and R. Seiler, Bounds for the adiabatic approximation with applications to quantum computation, J. Math. Phys. \textbf{48}, 102111 (2007), \href{https://arxiv.org/abs/quant-ph/0603175}{arXiv:quant-ph/0603175}.
\end{thebibliography}
\end{document}